# Combining Brillouin spectroscopy and machine learned interatomic potentials to probe mechanical properties of metal organic frameworks.

*Florian P. Lindner[1,2],* Nina Strasser[1], Martin Schultze[2], Sandro Wieser[3], Christian Slugovc[4], Kareem Elsayad[5], Kristie J. Koski[6], Egbert Zojer[1], Caterina Czibula[7]*

[1]Institute of Solid-State Physics, Graz University of Technology, Petersgasse 16, 8010 Graz, Austria

[2]Institute of Experimental Physics, Graz University of Technology, Petersgasse 16, 8010 Graz, Austria

[3]Institute of Materials Chemistry, Vienna University of Technology, Getreidemarkt 9, 1060 Wien, Austria

[4]Institute for Chemistry and Technology of Materials, Graz University of Technology, Stremayrgasse 9, 8010 Graz, Austria

[5]Division of Anatomy, Center for Anatomy and Cell Biology, Medical University of Vienna, Währinger Straße 13, 1090 Vienna, Austria

[6]Department of Chemistry, University of California Davis, 1 Shields Ave. 222 Chemistry, Davis CA, 95616, USA

[7]Institute of Bioproducts and Paper Technology, Graz University of Technology, Inffeldgasse 23, 8010 Graz, Austria

AUTHOR INFORMATION

**Corresponding Author**

*caterina.czibula@tugraz.at

*florian.lindner@tugraz.at

ABSTRACT The mechanical properties of metal-organic frameworks (MOFs) are of high fundamental and also practical relevance. A particularly intriguing technique for determining anisotropic elastic tensors is Brillouin scattering, which so far has rarely been used for highly complex materials like MOFs. In the present contribution, we apply this technique to study a newly synthesized MOF-type material, referred to as GUT2. We show that when combining the experiments with state-of-the-art simulations of elastic properties and phonon bands (based on machine-learned force fields and dispersion-corrected density-functional theory). This provides a comprehensive understanding of the experimental signals, which are correlated with the longitudinal and transverse sound velocities. Moreover, even when dealing with comparably small single crystals, which limit the range of accessible experimental data, combining the insights from simulations and experiments allows the determination of approximate values for the components of the elastic tensor of the studied material.

**TOC GRAPHICS**

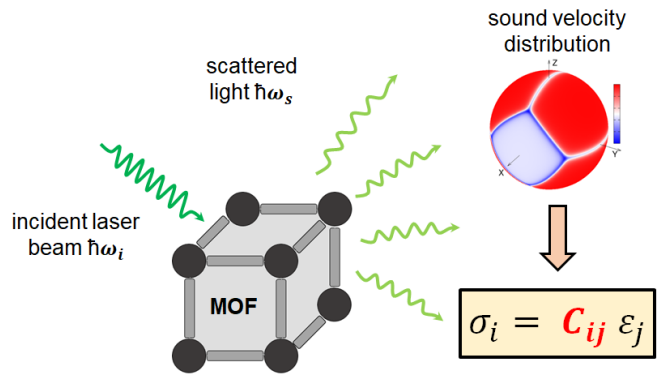

Almost 30 years after their discovery,[1,2] research interest in metal organic frameworks (MOFs) is still growing steadily. The versatile combination of inorganic and organic building blocks intrinsic to MOFs allows the formation of truly fascinating, aesthetically appealing, microporous, and complex but still crystalline structures.[3] A major reason for the lasting research interest in these materials is the plethora of their possible applications in various fields including gas storage, gas separation, catalysis, sensorics, and energy storage.[4,5,6,7] Aside from their functional properties, for virtually any of the envisaged applications the mechanical characteristics of the used MOFs are relevant. While for a gas storage application the available pore volume determines the hypothetical performance, for a real-world implementation one also needs to consider, how easily the MOF is deformed by mechanical forces during loading or de-loading cycles of the pores. High mechanical forces exerted on MOF crystals during operation could lead to a loss of structural integrity and in extreme cases even to the onset of amorphization.[8–10] It is, thus, evident that structural deformations under mechanical stress, which are typically quantified by engineering constants like Young modulus, E, or the Poisson's ratio, ν, will crucially affect the functional properties of MOFs. Hence, for many applications, a sound understanding of mechanical properties of MOFs is highly relevant. Despite their undeniable importance, and theoretical studies predicting mechanical properties MOFs,[11–14] literature on experimental data, especially on the single crystal level, to verify theoretical predictions, for example the elastic constants $C_{ij}$ of MOFs, is scarce. Besides pressure-dependent (powder) X-ray diffraction (PXRD) experiments,[15] elastic properties of MOFs were measured by nano-indentation or atomic force microscopy.[16] These methods are, however, prone to errors, caused by anisotropy of the probed sample and non-unidirectional stress fields generated by the indenter tip.[17–20] Measurements of bulk moduli in pressure PXRD experiments tend to give better agreement with theoretical predictions[21], but are often restricted to applying hydrostatic pressure, e.g., when using diamond-anvil cells.

In this letter, we demonstrate that Brillouin spectroscopy in combination with atomistic simulations is a promising, non-invasive experimental method to investigate the anisotropic mechanical properties of MOF single crystals in a contactless manner. It relies on analyzing light scattered from thermally excited acoustic phonons causing density fluctuations in the probed sample.[22–24] This provides access to the sound velocity tensor, from which the full elastic tensor, $C_{ij}$, of the studied material can be derived [25–28]. In the context of framework materials, Brillouin spectroscopy has, for example, been applied to study the mechanical properties of single crystals of the prototypical zeolitic imidazolate framework ZIF-8[29,30] and, recently, also perovskite-like dense MOFs[31].

Here, we conducted Brillouin experiments on a single crystal of a newly designed Zinc (II) MOF/coordination polymer, with chemical formula $C_{14}H_{18}N_4O_4Zn-H_2O$, which in the following will be referred to as GUT2 to be consistent with literature. Its synthesis and structural, thermal, and vibrational properties are described in Kodolitsch et al.[32] The material features the common characteristics of MOFs[33], consisting of metal-ion derived secondary building units connected by organic linker and displaying a non-negligible porosity. It crystalizes in the orthorhombic space group Pcca and contains 8 $Zn^{2+}$ ions in the primitive unit cell. Its density is calculated to be 1.502 g/cm³. The unit cell and crystal structure were determined by single crystal XRD (sc-XRD) measurements, which yielded lattice constants of: a = 15.1861(12) Å, b = 15.0082(13) Å and c = 15.0568(13) Å.[32] The crystal structure of the investigated MOF is visualized in Figure 1. It shows zinc ions tetrahedrally coordinated with two imidazole nitrogen atoms and two carboxylate groups. Two zinc ions are linked by two bridging ligands (2-methyl-imidazol) pointing in the same direction. Overall, the coordination of each zinc ion to two imidazole-nitrogen atoms and to two carboxylate ligands results in the formation of chains of coordination polymers. These are connected hydrogen bonds between the carboxylate groups. The interchain-bonding is reinforced by H-bonds involving water molecule in well-defined positions. The water molecules block the pore channels along the b-axis, see Figure 1, while the pore channels along the a- and c-axes remain open. Further details are provided in Kodolitsch et al.[32]

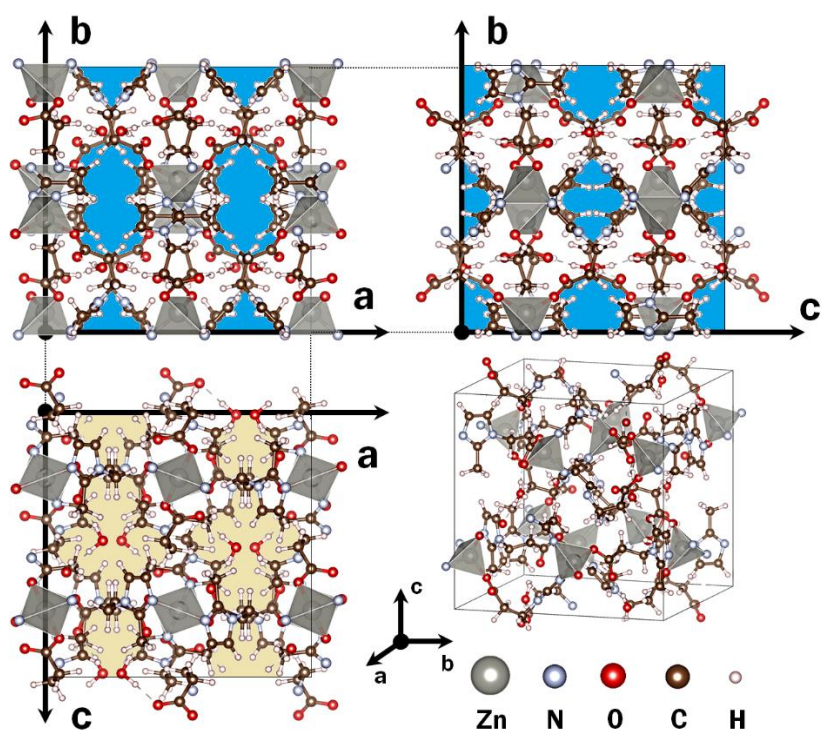

*Figure 1* Crystal structure of the studied zinc (II) coordination polymer GUT2 viewed from different directions. Pores blocked by adsorbed water molecules are shaded in yellow, whereas open pore channels along a- and c-directions are shaded in blue. Zinc coordination polyhedra are shaded in grey. The primitive unit cell is indicated by the thin black lines.

To measure the single crystal elastic constants of GUT2, a rhombohedrally shaped GUT2 crystal with an extent of roughly 0.5 mm was isolated and fixed by double sided adhesive tape onto a stainless-steel plate containing a 5 mm hole in its center, as shown in Figure 2. Subsequently, the stainless-steel plate holding the MOF crystal was attached to an optic rotation mount (Newport RSP-1T) to allow a precise adjustment of the sample rotation angle θ, as defined in Figure 2. Then the Brillouin scattering experiments were conducted in a 90° scattering geometry.[27,34] A schematic of this scattering geometry is shown in Figure 2. For details regarding the experimental setup (see the Supporting Information). In this scattering geometry, the wave vector **q** of the thermally activated phonons participating in the Brillouin scattering, is always within the scattering plane. The latter is defined by the wave vectors **k$_s$** and **k$_i$** of the scattered and incident laser light (532 nm). The collected scattered light was analyzed using a six-pass tandem Fabry-Perot interferometer (JRS Scientific Instruments, TFP-1). Starting from an arbitrary reference position for which the tilt angle was defined to be 0°, the sample was rotated in steps of 10° up to a tilt angle θ=90°. At tilt angles >90° the sample quality did not allow the detection of a high-quality Brillouin spectrum. The Brillouin shifts of the recorded spectra were obtained by fitting Lorentzian functions to the observable quasi longitudinal (QL) and quasi transversal (QT) peaks.

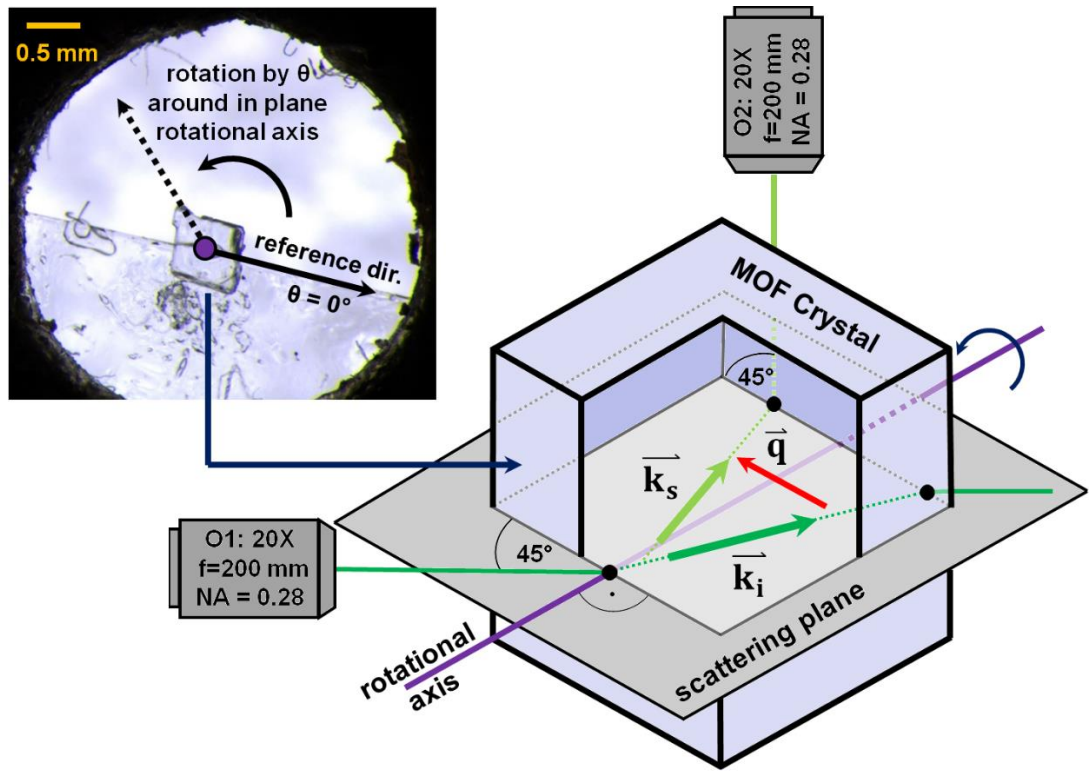

*Figure 2.* 90a Brillouin scattering geometry and single crystal MOF mounted on a metallic sample holder (top left). The optical paths of the incident (532 nm) and scattered laser light are shown in dark and light green respectively. Vectors $k_i$ and $k_s$ denote the wave vectors of incident- and scattered laser light. The wave vector of the thermally activated phonon involved in the scattering process is denoted by $q$. Refraction of the light at the surfaces of the sample is accounted for in the schematic sketch. By rotating the sample around the rotational axis (indicated in purple) by an angle $\theta$, directionally dependent sound velocities can be obtained.

A key challenge encountered in the studies of GUT2 is that the available MOF single crystals reached sizes of only a few 100 µm. This makes the handling of the investigated specimens during experiments difficult. Here, a tight integration of the experiments with state-of-the-art atomistic simulations of the elastic properties of GUT2 provides the necessary complementary information: on the one hand, the simulations ease the interpretation of the experimental results, and, on the other hand, they also justify approximations concerning the crystal symmetry, which will be made below. As described in the Supporting Information, a direct calculation of the elastic tensor elements $C_{ij}$ using density functional theory with fully converged plane-wave basis sets is hardly possible due to the complexity of the MOF material. Moreover, approximate approaches like the clamped-ion method implemented in the internal routines of, e.g., the VASP code[35–37] is numerically not stable here (see the Supporting Information). Thus, we resorted to the use of machine-learned potentials of the moment-tensor (MTP)[38] type parametrized following the procedure described by Wieser et al.[39] This involves combining VASP active learning[40] with an MTP parametrization utilizing the MLIP package[41] and yields potentials with essentially DFT accuracy[38] that are many orders of magnitude faster than the parent DFT approach. The availability of the MTPs allows a full relaxation of atomic positions in the strained cells and also the simulation of phonon band-structures employing converged GUT2 supercells. To confirm the appropriateness of the force-field calculated elastic constants, DFT calculations with a potentially less complete, but numerically more efficient basis sets were performed employing the CRYSTAL23 package [42,43] (see the Supporting Information). A comprehensive description of the applied simulation methodology can be found in the SI. While the simulations performed with the machine-learned potential in the following will be referred to as MTP results, the CRYSTAL23 results will be denoted DFT results. A typical spectrum observed in our Brillouin experiments is shown in Figure 3a. It displays two distinct peaks on either side of the elastic peak at 0 GHz. The two sets of peaks are shifted by ∼±5 GHz and by ∼±10 GHz. Their nature can be identified based on the calculated low-frequency phonon band structure of GUT2, displayed in Figure 3b. It reveals two nearly degenerate acoustic bands at lower frequencies and a single, higher energy acoustic band. Analyzing the degree of longitudinality of the bands shows that the higher band is of primarily longitudinal character, while the lower two bands are primarily transverse in nature (for details see the Supporting Information). These results allow identifying the peaks shifted by ~±10 GHz as quasi longitudinal (QL) and the ones shifted by ~±5 GHz as quasi transverse (QT). The near degeneracy of the transverse bands also explains, why in the measured Brillouin spectra depicted in Figure 3a, only one QT peak is resolved.

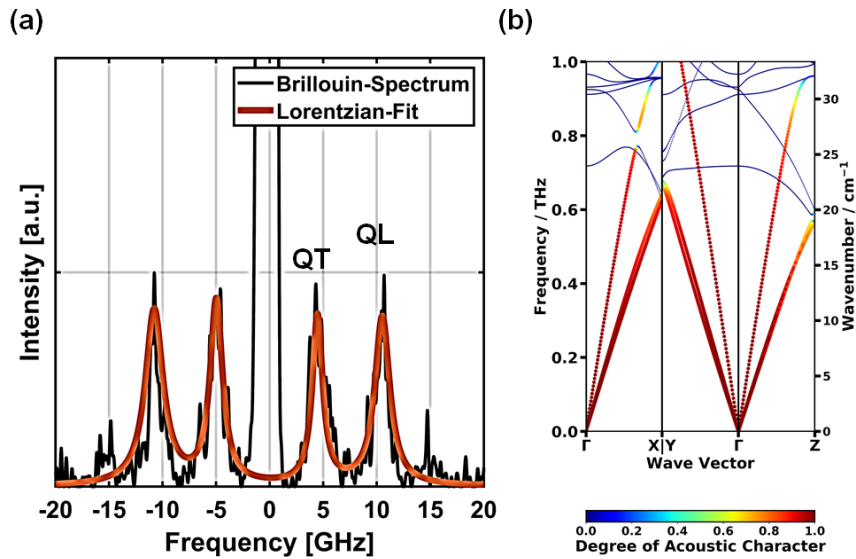

*Figure 3* *(a) Typical Brillouin spectrum of GUT2 with quasi longitudinal (QL) and quasi transversal (QT) peaks as observed during experiments (a). (b) MTP-calculated low-frequency phonon band structure of GUT2. The bands in (b) are colored according to their acoustic character[44].*

In the present scattering geometry, the velocity of the acoustic phonons, v, that were involved in the scattering process, is directly related to the observed frequency shift, $\Omega_B$, and the wave length, λ, (532 nm) of the incident laser light via: [45]

$$v = \frac{\lambda\, \Omega_B}{\sqrt{2}}$$

(1)

Using equation ( 1 ) , a distribution of sound velocities for different tilt angles θ, i.e., along different directions within the investigated MOF single crystal are obtained. These are shown in Figure 4.

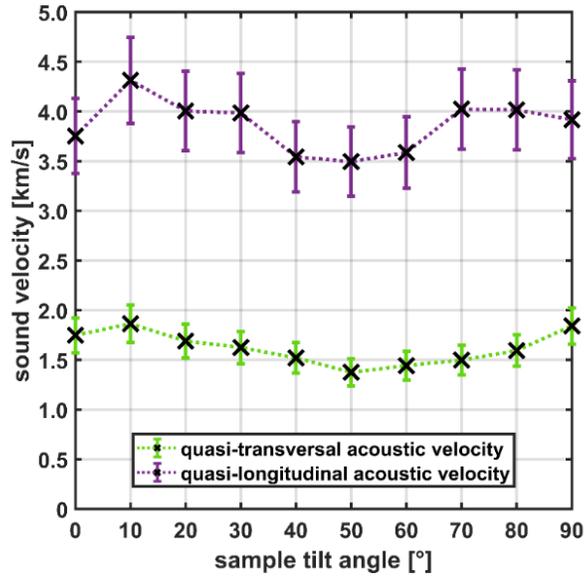

*Figure 4. Direction-dependent sound velocities in GUT2 derived from Brillouin experiments rotating the sample about the axis shown in Figure 1. The data points are drawn with an ad-hoc uncertainty bar of 10%.*

The data reveal that the transverse sound velocities in the MOF-2 crystal are within 1.4 km/s and 1.9 km/s, while the longitudinal sound speeds vary between 3.5 km/s and 4.3 km/s (see also Table 1). Importantly, the measured sound velocities agree rather well with the calculated ones, even though the spread of sound velocities is somewhat larger in the experiments (see Table 1). The situation in GUT2 is reminiscent of the observations for ZIF-8 by Tan et al..[29] There, however, smaller sound velocities between 1.0 km/s – 1.2 km/s for transverse and 3.1 km/s – 3.2 km/s for longitudinal waves were observed in similar measurements[29]. As will be detailed below, this suggests[45] that GUT2 is stiffer than ZIF-8. It is also interesting to compare the results for GUT2 with the sound velocity measured by phonon acoustic spectroscopy for the closed pore phase of the zinc-based, flexible MOF DUT-1(Zn). As in this case, v amounted to only 0.8 km/s, [46] one can conclude that even in its closed pore phase DUT-1(Zn) is significantly less stiff than both ZIF-8 and GUT2.

*Table 1.* *Measured and calculated speeds of sound, elastic constants $C_{ij}$ and Hill averaged mechanical properties for GUT2, extracted from Brillouin spectroscopy, and calculated using the moment tensor potential and DFT. The elastic constants are reported within the cubic approximation. Reported error bars are calculated assuming an ad hoc ±10% estimate for the uncertainties of the measured sound velocities.*

|  |  | Brillouin Spectroscopy | MTP | DFT |
|---|---|---|---|---|
|  | long. sound vel. (min / max) [km/s] | 3.49 ± 0.35 / 4.31 ± 0.43 | 3.77 / 4.20 | 3.54 / 3.93 |
|  | trans. sound vel. (min / max) [km/s] | 1.38 ± 0.14 / 1.86 ± 0.19 | 1.77 / 1.92 | 1.74 / 1.99 |
| cubic approximation | $C_{11}$ [GPa] | 18.4 ± 3.7 | 23.6 | 21.4 |
| cubic approximation | $C_{12}$ [GPa] | 12.7 ± 2.5 | 13.0 | 11.8 |
| cubic approximation | $C_{44}$ [GPa] | 5.2 ± 1.1 | 5.7 | 4.9 |
| cubic approximation | Bulk modulus K [GPa] | 14.6 ± 2.9 | 16.5 | 15.0 |
| cubic approximation | Young's modulus E [GPa] | 11.2 ± 2.3 | 14.9 | 13.1 |
| cubic approximation | Shear modulus G [GPa] | 4.1 ± 0.8 | 5.5 | 4.8 |
| cubic approximation | Poisson's ratio [1] | 0.372 ± 0.002 | 0.349 | 0.355 |

The fully quantitative determination of the components of a material's elastic tensor, $C_{ij}$, from measured sound velocities along specific crystallographic directions requires the inversion of the so-called Christoffel equation[47,48]. The latter describe the dispersion relation for plane sound waves traveling through a crystalline solid and relate direction-dependent sound velocities to a material's elastic constants $C_{ij}$[49]. A complication in the case of the orthorhombic GUT-2 arises from the fact that its elastic tensor contains nine independent elements. Despite considerable efforts the set of experimental datapoints is rather limited for the available single crystals. This prevents an unambiguous determination of all nine elements. Thus, to gain further insights, we pursued a dual approach: On the one hand, we extracted elastic constants assuming a higher crystal symmetry, which is motivated by the observation that the size and shape of the GUT2 unit cell is close to cubic with orthogonal lattice vectors, whose lengths vary by at most 1% (see above). Even more importantly, the calculated elastic tenor is reasonably close to cubic both for the MTP and DFT approaches. On the other hand, from the experiments we estimated limits to certain elements of the elastic tensor for the actual orthorhombic symmetry and compare these estimates to the calculated values. [41,50]

Within the cubic approximation, the calculated tensor elements $C_{ij}$ (listed in Table 1) are obtained by averaging over the respective elements of the orthorhombic elastic tensors, which are listed in the Supporting Information. Regarding the experiments, estimates for the elastic constants can be made based on the measured direction dependent values of the sound velocities for the longitudinal and transversal acoustic modes, using the so called 'envelope method'[30]. Originally introduced in high-pressure Brillouin experiments[50,51], the 'envelope method' allows deriving an estimate of the elastic constants of a crystalline sample with unknown orientation. It assumes that the measured direction-dependent sound velocities depicted in Figure 4, form an envelope to the maximum and minimum acoustic velocities, which, in the cubic case are straightforwardly related to the elements of the elastic tensor. The mathematical details of this approach for the case of GUT2 are described in the SI.[51] Notably, when applying the 'envelop' method to determine the elastic tensor of ZIF-8,[30] a very good agreement with the values obtained by a full inversion of Christoffel relations[29] was obtained. Assuming that the measured minimal quasi-longitudinal sound velocity (of 3.5 km/s) and the minima and maxima of the transverse sound velocity from Figure 4 (1.9 km/s and 1.4 km/s) are good estimates for the true extremum velocities, one then obtains the following estimates for the independent elements of the elastic tensor: $C_{11}$ = 18.4 GPa; $C_{44}$ = 5.2 GPa; $C_{12}$ = 12.7 GPa. As shown in Table 1, especially the values for $C_{44}$ and $C_{12}$ agree very well with both types of simulations. The value of $C_{11}$ extracted from the experiments is somewhat smaller than the calculated one, but the deviation is still acceptable.

To estimate, how well the sound velocity distributions are sampled in the experiments (c.f., Figure 4), we used the "christoffel" python package[49] in combination with the MTP- and DFT-calculated elastic tensors to solve the Christoffel dispersion relation in forward direction. The obtained simulated velocity distributions based on the MTP elastic tensor for the QL and QT modes is shown in *Figure 5* (with the DFT results in the Supporting Information). The calculated velocity distributions reveal that there is a reasonable correspondence between the ranges of the calculated and measured sound velocities, especially considering the experimental errors and the differences in the values obtained with the two simulation approaches (see Table 1). In fact, the observation that the experimental spread in sound velocities is even larger than the calculated one supports the assumption that the true range of sound velocities is suitably probed in the experiments.

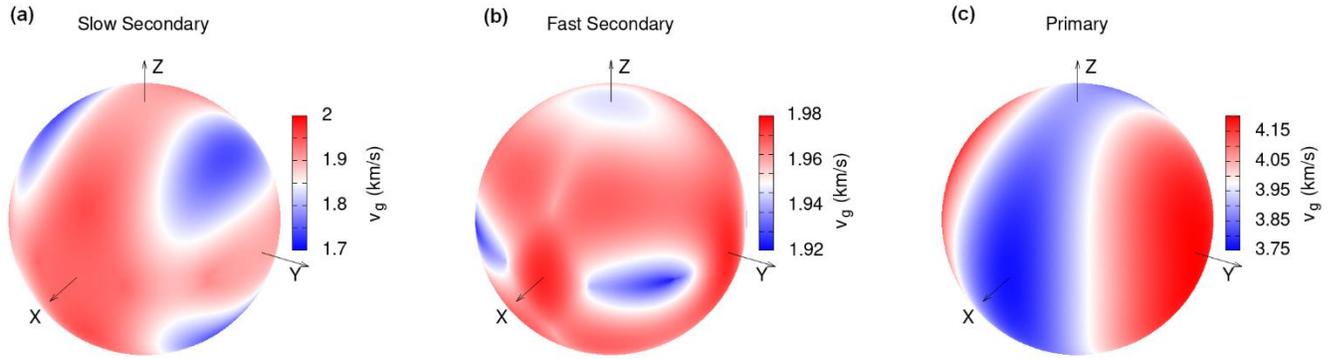

*Figure 5. Directional dependence of the MTP-based sound velocity distributions plotted on a unit sphere. (a) Slow quasi-transversal mode $v_{QT2}$, (b) fast quasi-transversal mode $v_{QT1}$ and (c) quasi-longitudinal mode $v_{QL}$. The similar DFT calculated distributions are contained in the Supporting Information.*

Notably, the directions of the maxima and minima of the sound velocities are consistent with the assumptions made in the 'envelop method' for a subset of the crystallographic directions (see discussion in the Supporting Information). Still, the angular distributions in Figure 5 do not display a cubic symmetry. Therefore, we also derived equations, which for the actual orthorhombic symmetry allow us to at least determine limits to the values of the components of the elastic tensor from the measured minimum and maximum sound velocities. Applying these relations (for details see SI), we find that the first three diagonal elements of the elastic tensor ($C_{11}$, $C_{22}$, and $C_{33}$) should be in the range between 18.5 GPa and 29.7 GPa, while $C_{44}$, $C_{55}$, and $C_{66}$ must be smaller than 5.2 GPa, which is consistent with the results presented above for the cubic approximation. For the set of diagonal elements related to compressive strain ($C_{11}$, $C_{22}$, and $C_{33}$), the conditions derived from the experiments are fulfilled for both, the MTP and the DFT simulated elastic tensor, while for the shear related components ($C_{11}$, $C_{22}$, and $C_{33}$) only the DFT results are strictly below the experimentally set limit, while the MTP results are slightly higher (by at most 0.6 GPa; see the Supporting Information).

From the elements of the elastic tensor, so-called engineering constants can be obtained. They are typically used to quantify the mechanical robustness of a material in technical applications and are derived from the components of the elastic tensor, using averaging schemes[52,53]. Averaged engineering constants like Young's, bulk or shear moduli are typically scalar values that account for the fact that in applications, usually one has to deal with polycrystalline samples, which can be assumed to behave in good approximation as isotropic. Using the ELATE package[54], we calculated the Hill averaged[53] Young's modulus E, shear modulus G, bulk modulus K and Poisson's ratio $\nu$ for GUT2. This was done using the Brillouin scattering elastic constants as well as the values from the MTP and DFT simulations in the cubic approximation (see Table 1). Again, the engineering constants derived from the simulations and from the experiments are in good agreement. The deviations are slightly larger between experiments and the MTP simulations, which we attribute to the larger (~20 %) deviation between the experimental and the MTP calculated values of $C_{11}$. In passing we mention that the values obtained for the averaged engineering constants when using the simulated, full orthorhombic elastic tensor are essentially identical to the results from the cubic approximation (see the Supporting Information).

To put the obtained results into perspective, we compare the elastic constants of GUT2 to those obtained in the past for two prototypical cubic systems, ZIF-8 and MOF-5: comparing the values reported by Tan et al[29] for ZIF-8 ($C_{11}$ = 9.5 GPa, $C_{12}$ = 6.9 GPa, and $C_{44}$ = 0.9 GPa) with the elastic constants of GUT2, one sees that ZIF-8 displays a significantly larger structural flexibility. The most striking difference is that the shear constant $C_{44}$ of GUT2 is almost 6 times as large as the one reported for ZIF-8. The above trends prevail for the engineering constants: Tan et al. report a Hill average for the shear modulus of ~1.1 GPa for ZIF-8, which suggests an almost 4 times lower resistance against shear stresses in ZIF-8 than in GUT2. Furthermore, GUT2 with its bulk modulus of 14.6 GPa and a Young's modulus of 11.2 GPa is almost twice as resistant against hydrostatic compression and uniaxial loading as ZIF-8 in which the latter were reported to be 7.7 GPa and 3.1 GPa respectively[29].

For MOF-5 (as the prototypical example of an isoreticular MOF), to the best of our knowledge, no experimental values of the elastic constants are available. Nevertheless, it is interesting to compare our results for GUT2 to the available theoretical predictions for MOF-5: based on GGA level DFT calculations, Bahr et al.[55] report the elastic constants for MOF-5 to be $C_{11}$ = 27.8 GPa, $C_{12}$ = 10.6 GPa and $C_{44}$ = 3.6 GPa. This means that for this classical isoreticular MOF rather similar elastic constants as in GUT2 are to be expected.

In summary, we conducted state-of-the-art Brillouin scattering experiments, for a newly synthesized MOF and deduced its mechanical properties from single crystal elastic constants $C_{ij}$. These were obtained from measured direction dependent sound velocities. Using a machine learned moment tensor potential as well as within DFT, we were able to also simulate the elements of the MOF's elastic tensor. The main challenge faced in the evaluation of the experimental data was the rather complex nature of the elastic tensor of GUT2 resulting from its the non-cubic (i.e., orthorhombic) unit. To overcome this issue, the crystal symmetry in the data evaluation was approximated as cubic and, as an alternative approach, estimates for certain tensor elements of the true, orthorhombic system were performed. Despite these challenges, an overall very good agreement between the elastic constants derived from the experiments and from the simulations was obtained. This also testifies to the predictive power of the numerically extremely efficient machine learned potential applied here. Thus, in future studies this type of potential will be used in combination with molecular dynamics simulations to include the effects of temperature or of external loading in the prediction of the mechanical properties of MOFs. This will also lead to an even tighter connection between theory and experiments, with the latter typically performed at elevated temperatures. This should provide an avenue for obtaining the elastic constants of MOFs in a reliable and efficient manner, complementing established methods like pressure-dependent powder X-ray diffraction or Raman spectroscopy.